\documentclass[12pt]{iopart}
\usepackage{iopams}
\usepackage{graphicx}
\usepackage{xcolor}
\usepackage{float}
\begin{document}
\title[]{First-order phase transformation at constant volume: a continuous transition?}

\author{V. F. Correa \& F. J. Castro}

\address{Centro At\'omico Bariloche (CNEA) and Instituto Balseiro (U. N. Cuyo), 8400 Bariloche, R\'io Negro, Argentina}

\ead{victor.correa@cab.cnea.gov.ar, fcastro@cab.cnea.gov.ar}

\vspace{10pt}
\begin{indented}
\item[] July 2021
\end{indented}

\begin{abstract}
We describe a first-order phase transition of a simple system in a process where the volume is kept constant. We show that, unlike what happens when the pressure is constant, (i) the transformation extends over a finite temperature (and pressure) range, (ii) each and every extensive potential (internal energy $U$, enthalpy $H$, Helmholtz energy $F$ and Gibbs energy $G$), and the entropy $S$, is continuous across the transition, and (iii) the constant-volume heat capacity does not diverge during the transition, only exhibits discrete jumps.
These non-intuitive results highlight the importance of controlling the correct variables in order to distinguish between continuous and discontinuous transitions. Additionally, they provide a didactic tool to further discuss the phase transitions phenomena.
We apply our results to describe the transition between ice VI and liquid water using thermodynamic information available in the literature.
\end{abstract}

\vspace{2pc}

\submitto{\EJP}
\maketitle

\normalsize

\section{Introduction}

Phase transitions (PT) are probably one of the most interesting and conceptually-rich phenomena approached by Thermodynamics and/or Statistical Mechanics. 
The classical traditional classification presented in 1933 by P. Ehrenfest \cite{Ehrenfest1933} introduces the concept of transition order: when at least one of the first order derivatives of the Gibbs energy $G(T,p)$ with respect to its natural variables, temperature $T$ and pressure $p$, shows a  jump discontinuity, the transition is said to be first order; if all the first-order derivatives are continuous but at least one of the second-order derivatives shows a jump discontinuity, the transition is said to be second order; and so on for higher-than-second-order transitions. 
Since then, this scheme has become universally accepted due to its simplicity and conceptual content. Another advantage of this classification is that it can be rapidly understood and assimilated by students at the undergraduate level.
Even after the development of modern critical phenomena theories and related concepts like order parameter, correlation length, fluctuations and symmetry, the classification has remained valid in a simplified form: first-order or discontinuous transitions on one side, and continuous transitions on the other \cite{Jaeger1998}. 

Though the classification was originally thought for a simple system characterized by the variables $T$, $p$ and the volume $V$ (of which only two are independent), it can be generalized to include other variables such as electric and magnetic fields, or strain-stress effects, as long as the system remains thermodynamic: large enough as to neglect surface and geometrical effects, and without long-range interactions that could invalidate the additive nature of the extensive variables.

Typical questions that often arise from students regarding this Ehrenfest scheme are: why is it formulated in terms of $G(T,p)$? What makes this thermodynamic potential so special?
From the experimental point of view, the first studies on PT made at constant pressure and temperature realized the presence of a latent heat $L$ and/or a volume change $\Delta V$ across the transition. In fact, it was the lack of any observable $L$ or $\Delta V$ in the at that time newly discovered superfluid transition that triggered the Ehrenfest work \cite{Jaeger1998}. As the latent heat L can be connected with the entropy $S$ ($L = T_0 \Delta S$, where $T_0$ is the transition temperature), and entropy and volume $V$ are first order derivatives of the Gibbs energy, it seems natural to give a classification in terms of $G(T,p)$. 

Besides this, there are additional reasons that point towards $G(T,p)$ as a special thermodynamic potential to analyze PT. These reasons naturally arise when examining phase transitions under different experimental conditions. A first scenario to do this is constant volume and temperature, but we will show in the following  that a textbook first-order transition, with finite $L$ and $\Delta V$, in a simple system, cannot take place under constant $V$ and $T$ conditions. A second scenario is constant volume only. We will examine this process in detail and demonstrate that: (i) the transformation extends over a finite range of $T$ (and $p$), and (ii) each and every extensive potential (internal energy $U$, enthalpy $H$, Helmholtz energy $F$ and Gibbs energy $G$) and the entropy $S$ is continuous across the transition when $V$ is constant. 
Additionally we will discuss the behavior of the heat capacity and constant volume $C_V$ during the transition and show that, contrary to what happens to the constant pressure heat capacity $C_p$ during a constant $T$ and $p$ phase transition, there is no divergence in $C_V$, only discrete jumps. In the final section of the paper the results are illustrated by analyzing a real PT: the transition between liquid water and solid ice-VI under constant volume.

\section{Statement of the problem and equilibrium condition}

\subsection{Constant pressure transformation}
\label{sub1}

\begin{figure}[t]
\centering
\includegraphics[scale=0.7]{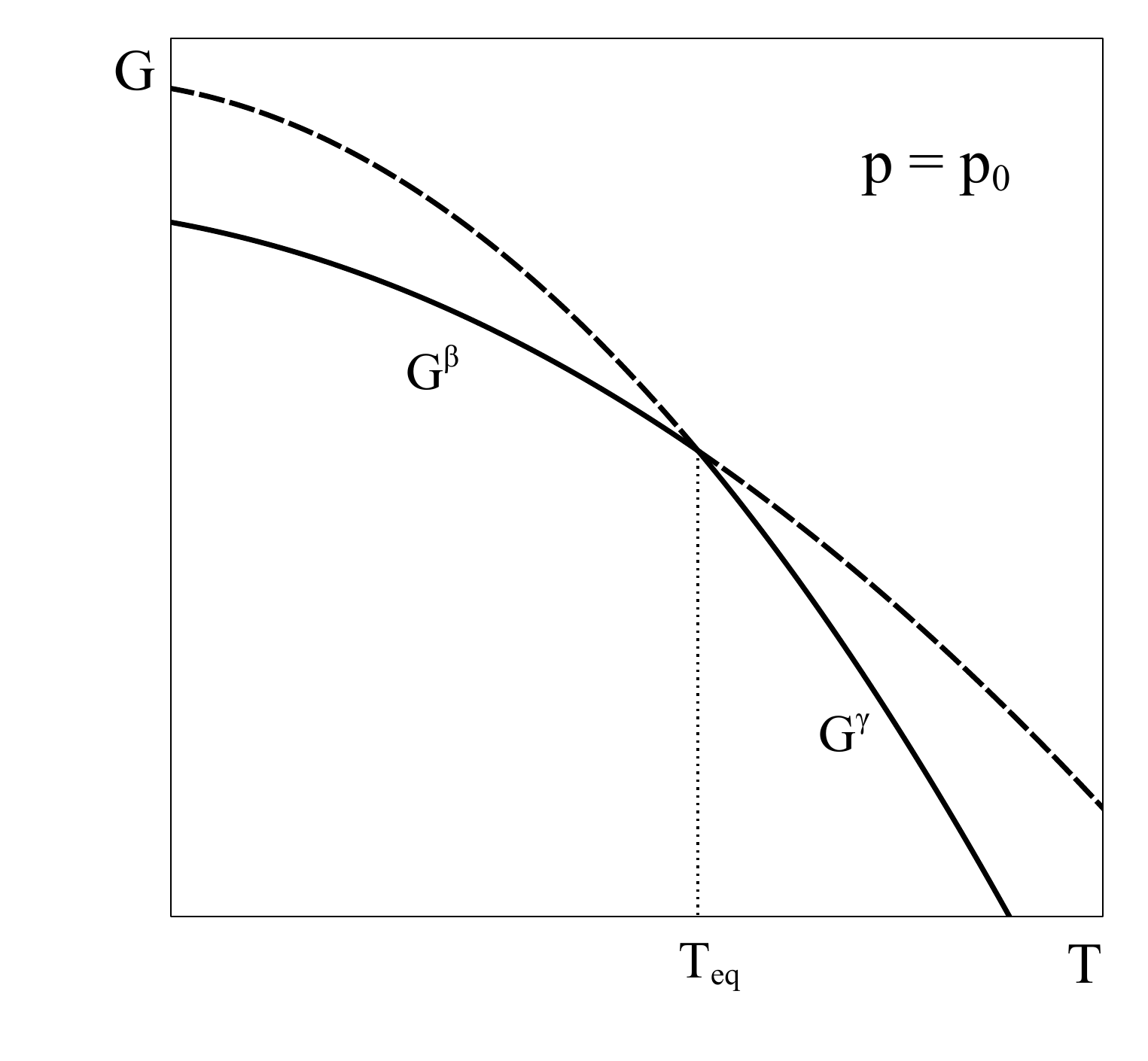}
\caption[]{Qualitative Gibbs energy curves of phases $\beta$ and $\gamma$ as a function of temperature at a constant pressure $p_0$. Both phases coexist in equilibrium at $T=T_{eq}$. Solid (dash) lines correspond to stable (metastable) states.}
\label{fig1}
\end{figure}

Let us begin by recalling phase equilibrium and the characteristics of a first-order transition at constant $p=p_0$. Consider a simple system with a well defined composition, and completely characterized by the variables $p$, $T$ and $V(T,p)$. This macroscopic system can exist in two different phases ($\beta$ and $\gamma$) in the domain of interest of the $p-T$ phase diagram, and the transformation between these phases is a first-order transformation, with finite enthalpy and volume changes associated with it. The classical analysis of phase equilibrium in this case is typically made by plotting the characteristic Gibbs functions of phases $\beta$ and $\gamma$ as a function of temperature at the constant pressure $p_0$ as shown in Fig. \ref{fig1}. 

At each temperature $T$, the state of equilibrium of the system corresponds to the phase that has the lower Gibbs energy, as equilibrium under constant $T$ and $p$ corresponds to the state that minimizes the Gibbs energy. Therefore, for temperatures below $T_{eq}$, $\beta$ is the stable phase, and above $T_{eq}$, $\gamma$ is the stable phase. At $T=T_{eq}$ the Gibbs energy curves coincide, and phases $\beta$ and $\gamma$ can coexist in equilibrium (Fig. \ref{fig1}). Hence, if the system is slowly heated at constant pressure $p_0$ starting in an equilibrium state in the $\beta$ single phase field at temperature $T_0$, it keeps in this phase until temperature $T_{eq}$ is reached. At this temperature phase transformation under equilibrium begins, and the system gradually moves from $\beta$ phase to $\gamma$ phase. During this transformation, heat is evolved according to the enthalpy difference between phases $\beta$ and $\gamma$, and the volume of the system changes following the differences of molar volume between $\beta$ and $\gamma$ phases. The finite change of slope in passing from the Gibbs energy of $\beta$ phase to that of $\gamma$ corresponds to the finite entropy change characteristic of this first-order phase transformation.

The amount of each phase during the transformation can be described by a single parameter, the extent of transformation $\xi$, defined as

\begin{equation} 
n^\beta = 1-\xi
\qquad \qquad
n^\gamma = \xi
\label{eqn1}
\end{equation} 

\noindent where $n^\beta$ and $n^\gamma$ are the number of moles of $\beta$ and $\gamma$, respectively, and for simplicity we are considering 1 mole of substance. Every extensive quantity of the complete system can be expressed using the extent of transformation and the corresponding molar quantity. For example, the system volume and enthalpy are given by:

\begin{eqnarray} 
V(T,p) &=& V_m^{\beta}(T,p) \, n^{\beta} + V_m^{\gamma}(T,p) \, n^{\gamma}  \label{eqnF2a} \nonumber \\
\nonumber \\
&=& V_m^\beta(T,p) + \left[ V_m^\gamma(T,p)-V_m^\beta(T,p) \right] \, \xi, \nonumber \\
\nonumber \\
&=& V_m^\beta(T,p) + \Delta V_m^{PT} (T,p) \xi.
\label{eqn2}
\end{eqnarray} 

\begin{eqnarray} 
H(T,p) &=& H_m^{\beta}(T,p) \, n^{\beta} + H_m^{\gamma}(T,p) \, n^{\gamma}  \nonumber \\
\nonumber \\
&=& H_m^\beta(T,p) + \left[ H_m^\gamma(T,p)-H_m^\beta(T,p) \right] \, \xi, \nonumber \\
\nonumber \\
&=& H_m^\beta(T,p) + \Delta H_m^{PT} (T,p) \xi. \label{eqn3}\\
\nonumber\
\end{eqnarray} 

\noindent Where $\Delta V_m^{PT} (T,p)$ and $\Delta H_m^{PT} (T,p)=L$ denote the volume and enthalpy changes (latent heat) associated with the phase transformation, respectively. In fact, by measuring the heat taken or released by the system (or the volume) during the phase transformation, the value of $\xi$ can be inferred, and from it, the value of any extensive quantity of the system can be calculated by using expressions similar to equations (\ref{eqn2}) and (\ref{eqn3}).

If the enthalpy of the complete system is analyzed as a function of temperature, at $T=T_{eq}$ the characteristic jump associated with the latent heat is found. It is interesting to note that the jump or discontinuity is associated with the complete phase transformation. During the transformation, continuous values of enthalpy can be attributed to the system at $T=T_{eq}$  by means of the extent of transformation $\xi$ following eq. (\ref{eqn3}). The temperature derivative of this curve corresponds to the constant pressure heat capacity of the system (defined by $\left. C_p = \frac{\partial H}{\partial T} \right)_{p} = T \left. \frac{\partial S}{\partial T} \right)_{p}$). The limits of this expression reaching the transition point from the left or from the right are different, because they correspond to the temperature derivative of the enthalpy of $\beta$ or $\gamma$ phase, respectively. Additionally, during the transition, there is a finite enthalpy change with no associated temperature change. Therefore, there is a divergence in $C_P$ that in an ideal representation of an equilibrium phase transformation can be described by a Dirac $\delta$-function. A similar reasoning shows that the volume thermal-expansion coefficient $\alpha = \frac 1 V \left. \frac{\partial V}{\partial T} \right)_{p}$ and the isothermal compressibility $\kappa = -\frac 1 V \left. \frac{\partial V}{\partial p} \right)_{T}$ exhibit similar behavior during a constant pressure transition.

\subsection{Constant volume transformation}
\label{Vcte}

\begin{figure}[t]
\centering
\includegraphics[scale=0.7]{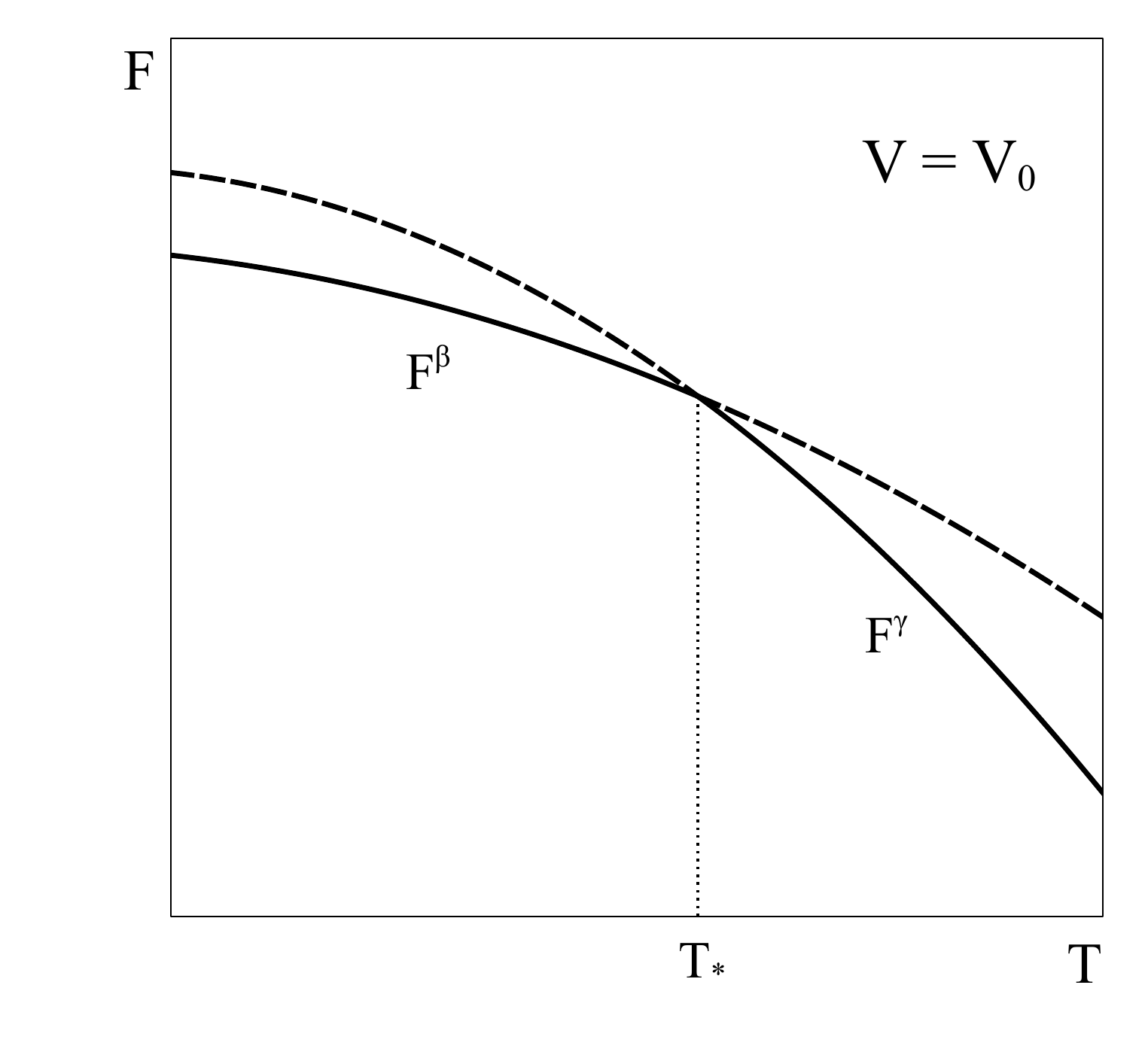}
\caption[]{Qualitative Helmholtz energy curves of phases $\beta$ and $\gamma$ as a function of temperature at a constant volume $V_0$.}
\label{fig2}
\end{figure}

Let us now consider an approach similar to the previous one, but applied to a constant volume transition. As the equilibrium of the system under constant $T$ and $V$ is given by the minimum of the Helmholtz potential, let us schematically represent typical Helmholtz curves of phases $\beta$ and $\gamma$ as a function of temperature under constant $V=V_0$ conditions (Fig. \ref{fig2}). 

By following a similar reasoning to that of subsection \ref{sub1} it could be concluded that phase $\beta$ is the stable phase below temperature $T_*$ and phase $\gamma$ is the stable one above it. Additionally, it could be thought that at $T_*$ both phases could coexist in equilibrium. But this argument has two flaws. First, nothing ensures that at volume $V_0$ and $T=T_*$ the pressure of phase $\beta$ given by its equation of state matches that of phase $\gamma$. In fact, in general this condition is not met. Therefore, the equality of all the characteristic intensive variables of each phase, a necessary requirement for the equilibrium of two phases, would not be satisfied \cite{Callen1985,Landau1969}. Secondly, nothing ensures that there is no other Helmholtz energy curve that fulfills the constant volume condition and also lies below both single-phase Helmholtz energy curves, so giving the actual minimum. This could be done, for example, by combining different amounts of phases $\beta$ and $\gamma$ at each temperature, thanks to the fact that volume is an extensive quantity. In fact, in the following section we will see that such a curve indeed exists! 

This analogy highlights another dimensions of the importance of Gibbs energies. First, by exclusively depending on the intensive variables $p$ and $T$, it can be assured that when two Gibbs energy curves at constant p intersect at an equilibrium temperature (or two curves at constant $T$ intersect at an equilibrium pressure) the other intensive variable also coincides, ensuring equilibrium. Secondly, the absence of natural extensive variables of $G$ precludes the possibility of combining two single phase Gibbs energy curves into a new curve that could lie below both single-phase characteristic G curves.

\section{Behavior of thermodynamic quantities during a constant volume phase transformation}

\begin{figure}[t]
\centering
\includegraphics[width=\columnwidth]{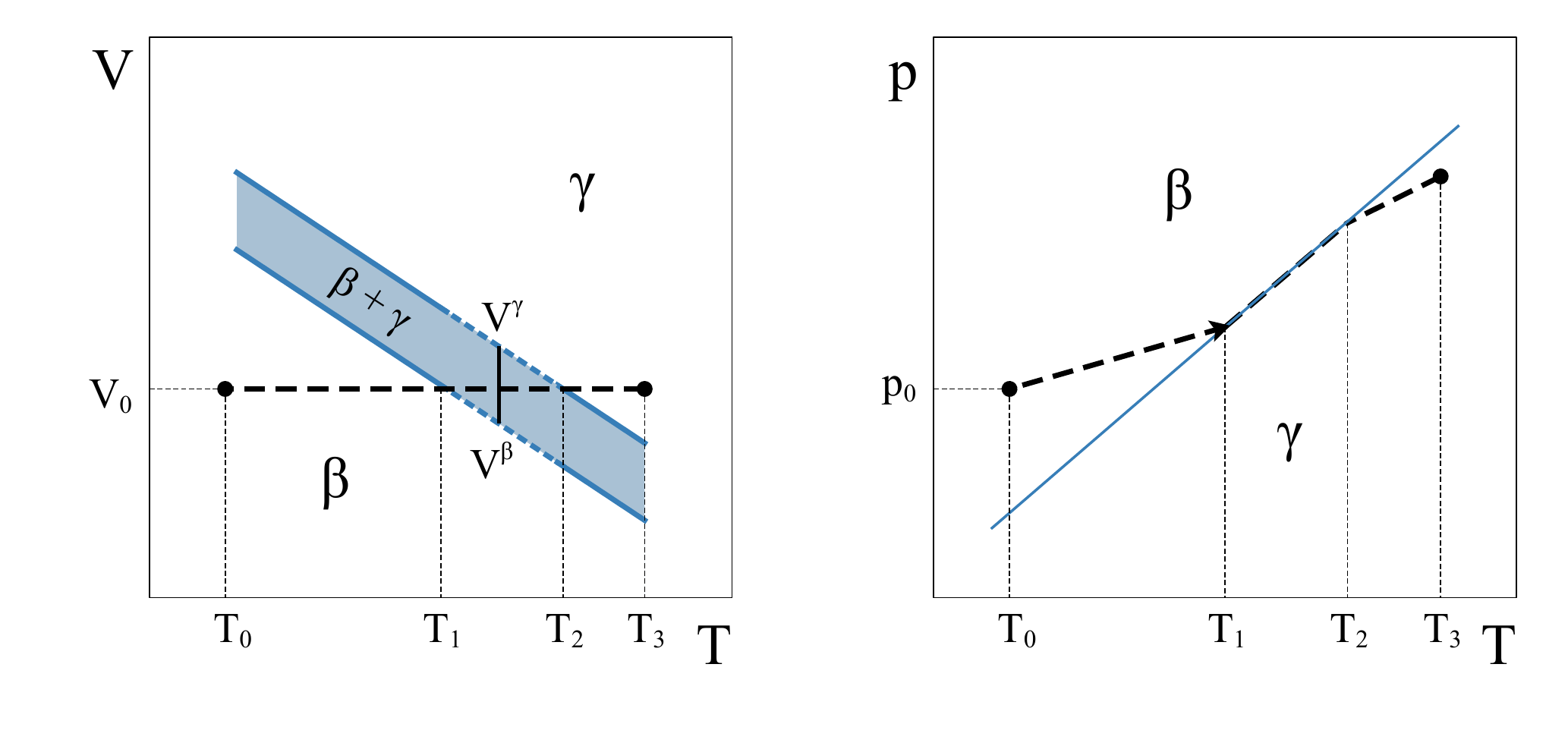}
\caption[]{Volume-Temperature (left) and Pressure-Temperature (right) schematic phase diagrams of a simple system. Two phases, $\beta$ and $\gamma$, transform each other via a first-order transition. The dash-line corresponds to a constant volume process starting at $V_0(T_0,p_0)$.}
\label{fig3}
\end{figure}

Therefore, how can a constant volume phase transformation be analyzed? The first thing to notice is that a constant volume first-order transformation must occur under a temperature range. To see this, let us go back to our simple system analyzed in Subsection \ref{sub1}. A $V-T$ phase diagram for this system (Fig. \ref{fig3}, left panel) shows two single phase regions and an area in the $V-T$ plane where $\beta$ and $\gamma$ phases coexist. The existence of an area instead of a line, is a consequence of the finite jump in volume at each temperature associated with the first order PT. If we want now to transform from phase $\beta$ to phase $\gamma$ at a constant volume $V_0$, the temperature during the phase transformation must necessarily change. By starting in the $\beta$ single phase field at $V_0(T_0,p_0)$ and warming up the system at constant volume, $\gamma$ will start nucleating at $T_1$, but the phase transformation will not end until temperature $T_2$ is reached (Fig. \ref{fig3}). During the coexistence both phases must be at equilibrium. This condition requires the equality of temperature, pressure and chemical potential \cite{Callen1985,Landau1969}. These conditions are met if the system evolves along the $p-T$ coexistence line in the $p-T$ phase diagram (Fig. \ref{fig3}, right panel). Therefore, during phase coexistence $p$ and $T$ are not independent, they are linked by the coexistence line, and the volume is fixed by the constant volume condition. Hence, there is only one degree of freedom. 

Algebraically the need for phase coexistence to occur over a finite range of temperatures can be seen by taking the differential of the system volume (\ref{eqn2}) and imposing the constant volume condition:

\begin{equation}
0=dV = dV_m^{\beta}(T,p) \, n^{\beta} + dV_m^{\gamma}(T,p) \, n^{\gamma} + V_m^{\beta}(T,p) \, dn^{\beta} + V_m^{\gamma}(T,p) \, dn^{\gamma}  \label{eqn4}
\end{equation}

\noindent Using the extent of transformation  $\xi$ (eq. \ref{eqn1}), the volume thermal-expansion coefficient $\alpha$ and the isothermal compressibility $\kappa$ for each phase, equation (\ref{eqn4}) can be rewritten as:

\begin{equation}
0= (\alpha^{\beta} V_m^{\beta} n^{\beta} + \alpha^{\gamma} V_m^{\gamma} n^{\gamma})dT - (\kappa^{\beta} V_m^{\beta} n^{\beta} + \kappa^{\gamma} V_m^{\gamma} n^{\gamma})dp + (V_m^{\gamma} - V_m^{\beta}) d\xi
\label{eqn5}
\end{equation}

\noindent where for readability the dependence of $\alpha$, $\kappa$ and the molar volumes on pressure and temperature has not been explicitly written. As we mentioned before, the equilibrium between $\beta$ and $\gamma$ phases during the transition makes temperature and pressure non-independent quantities. At each temperature the pressure is given by the coexistence line $p=p_c(T)$. Therefore, equation (\ref{eqn5}) can be rewritten as:

\begin{eqnarray}
0&=& \left[ \alpha^{\beta} V_m^{\beta} n^{\beta} + \alpha^{\gamma} V_m^{\gamma} n^{\gamma} - (\kappa^{\beta} V_m^{\beta} n^{\beta} + \kappa^{\gamma} V_m^{\gamma} n^{\gamma}) \frac {dp_C}{dT} \right] dT   \nonumber \\
\nonumber \\
& &+ (V_m^{\gamma} - V_m^{\beta}) d\xi
\label{eqn6}
\end{eqnarray}

\noindent We can see here that for the constant volume transformation to advance ($d\xi \neq 0$), the temperature (and pressure) must change ($dT \neq 0$). In other words, as the $\beta$ and $\gamma$ molar volumes are different, due to the first-order nature of the transition, the only way of keeping the volume constant during a phase transformation is by changing temperature and pressure in order to compensate the differences in molar volumes. 
In this sense, a first-order transformation at constant volume $V_0$ occurring at a single equilibrium temperature $T_*$ (as sketched in Fig. \ref{fig2}) is feasible if and only if the volume change across the transition vanishes, {\it i.e.} if $V^\beta(T_*)$ = $V^\gamma(T_*)$ = $V_0$. 

Therefore, we have proven that the transition at constant $V$ extends over a finite range of temperature and pressure, unlike what happens when the transformation occurs at constant $p$. 
During the transition the extent of reaction $\xi$ can be expressed as a function of $T$. By imposing the constant volume condition in equation (\ref{eqn2}) and taking into account that during the transition $p=p_C(T)$, $\xi$ can be expressed as:

\begin{equation}
\xi(T)=\frac{V_0-V_m^\beta[T, p_C(T)]}{V_m^{\gamma} [T,p_C(T)]-V_m^{\beta} [T,p_C(T)]}=\frac{V_0-V_m^\beta(T)}{\Delta V_m^{PT} (T)}
\label{eqn7}
\end{equation}

\noindent Two straightforward conclusions can be drawn from this equation: (i) $\xi(T_1) =$ 0, $\xi(T_2) =$ 1 since $V_m^{\beta}(T_1)=V_0$ and $V_m^{\gamma}(T_2)=V_0$, respectively; and (ii) $\xi(T)$ is a continuous function since $V_0$ is constant and the molar volumes of both phases are continuous functions.

Between $T_1$ and $T_2$, the mole numbers of each phase are explicitly given by 

\begin{equation} 
n^\beta(T) = \frac{V_m^\gamma(T) - V}{V_m^\gamma(T)-V_m^\beta(T)}
\qquad \qquad
n^\gamma(T) = \frac{V-V_m^\beta(T) }{V_m^\gamma(T)-V_m^\beta(T)} 
\label{eqn8}
\end{equation}

\subsection{Behavior of thermodynamic potentials}

Let us consider now a general extensive variable $Z$ and calculate its value during the coexistence at constant $V=V_0$. According to Euler's theorem, $Z$ can be calculated as

\begin{equation} 
Z(T) = Z \left[ p_c(T),T \right] = n^\beta \, Z_m^\beta \left[ p_c(T),T \right] +  n^\gamma \, Z_m^\gamma \left[ p_c(T),T \right]
\label{eqn9}
\end{equation}

\noindent which in terms of the extent of transformation reads

\begin{eqnarray} 
Z(T) &=& Z_m^\beta(T) + \left[ Z_m^\gamma(T)-Z_m^\beta(T) \right] \xi(T)   \nonumber \\
\nonumber \\
&=& Z_m^\beta(T) + \Delta_{PT} Z_m (T) \, \xi(T)
\label{eqn10}
\end{eqnarray}

\noindent Here $\Delta_{PT} Z_m (T) = Z_m^{\gamma}(T)-Z_m^{\beta}(T)$ represents the change of $Z_m$ across the transformation. If we take $Z(T_1) = Z_m^{\beta}(T_1)$ as a reference state and calculate the change in Z from this reference state. We obtain 

\begin{equation} 
\Delta Z(T) = Z(T) - Z_m^\beta(T_1) = Z_m^ \beta(T) - Z_m^\beta(T_1) + \Delta_{PT} Z_m (T) \, \xi(T)
\label{eqn11}
\end{equation}

\noindent We can identify two contributions to $\Delta Z(T)$: the change of $Z_m^{\beta}$ due to the variation of $T$ and the change due to the phase transformation itself. A quick overview of this equation allows us to see that $\Delta Z(T)$ is continuous as long as $Z_m^{i}$ ($i=\beta,\gamma$) is well defined and continuous in the whole temperature range, and this is indeed true for each and every extensive potential: $U, F, H, G$ and the entropy $S$. In other words, during a phase transformation at constant volume there are no jumps or discontinuities in any thermodynamic potential at any single temperature. This might seem rather puzzling at a first glance, as the transformation appears as a continuous one. But this appearance is nothing but the consequence of the transformation to extend over a finite temperature (and pressure) range. The characteristic jump of the first-order transition is spread over a temperature range when the transformation is done at constant volume, making it appear as if it were a continuous change.

We give next detailed expressions for the changes of all the thermodynamic potentials ($U, F, H, G$) and the entropy $S$ across a constant volume transition. The expressions can be explicitly calculated if the quantities $F_m^{i}(T,V)$, $V_m^{i}$ and $p_c(T)$ are known. 

Let us start with the Helmholtz free energy $F(T,V)$. The contribution from the transformation is

\begin{equation} 
\Delta_{PT} F_m (T) = \Delta_{PT} \left[ G_m (T)-p_c(T) V_m( T) \right] = -p_c(T)\Delta_{PT} V_m (T)
\label{eqn12}
\end{equation}

\noindent where we have used the equality of the chemical potentials during the transformation $G_m^{\beta} = G_m^{\gamma}$. Finally, we get

\begin{equation} 
\Delta F(T) = F_m^\beta(T) - F_m^\beta(T_1)-p_c(T) \Delta_{PT} V_m (T) \, \xi(T).
\label{eqn13}
\end{equation}

Proceeding with the entropy, from the Clapeyron equation we get

\begin{equation} 
\Delta_{PT} S_m (T) = \Delta_{PT} V_m (T) \frac{dp_c(T)}{dT} 
\label{eqn14}
\end{equation}

\noindent so the entropy change is 

\begin{eqnarray} 
\Delta S(T) &=& \left. \frac{\partial F_m^\beta}{\partial T} \right)_{V_m^{\beta}=V} (T_1) - \left. \frac{\partial F_m^\beta}{\partial T} \right)_{V_m^{\beta}=V} (T) \nonumber \\
\nonumber \\
&& + \Delta_{PT} V_m (T) \frac{dp_c(T)}{dT} \, \xi(T)
\label{eqn15}
\end{eqnarray}

\noindent It is easy to check that the same result is obtained by differentiation of equation (\ref{eqn13}), {\it i.e.} $\Delta S(T) = \left. \frac{\partial \Delta F}{\partial T} \right)_{V} - S_m^{\beta}(T_1)$.

As for the Gibbs energy, the change due to the transformation is null (equality of chemical potentials), thus the total change is

\begin{eqnarray} 
\Delta G(T) &=& G_m^\beta(T) - G_m^\beta(T_1)  \nonumber \\
\nonumber \\
&=& \left[ F_m^\beta(T) + p_c(T) \, V_m^\beta(T) \right] - \left[ F_m^\beta(T_1) + p_c(T_1) \, V_m^\beta(T_1) \right]  \nonumber \\
\nonumber \\
&=& F_m^\beta(T) - F_m^\beta(T_1) \nonumber \\
\nonumber \\
& & - \left. \frac{\partial F_m^\beta}{\partial V_m} \right)_T (T) \, V_m^\beta(T) + \left. \frac{\partial F_m^\beta}{\partial V_m} \right)_T ( T_1) \, V_m^\beta(T_1).
\label{eqn16}
\end{eqnarray} 

We can proceed similarly to obtain the internal energy 

\begin{eqnarray} 
\Delta U(T) &=& U_m^\beta(T) - U_m^\beta(T_1) + \Delta_{PT} U_m (T) \, \xi(T)  \nonumber \\
\nonumber \\
&=& \left[ F_m^\beta(T) + T S_m^\beta(T) \right] \nonumber \\
\nonumber \\
& & - \left[ F_m^\beta(T_1) + T_1 S_m^\beta(T_1) \right] + \left[ \Delta_{t} F_m (T) + T \Delta_{PT} S_m (T) \right] \, \xi(T) \nonumber \\
\nonumber \\
&=& F_m^\beta(T) - F_m^\beta(T_1)- T \left. \frac{\partial F_m^\beta}{\partial T} \right )_{V_m^{\beta}=V} (T) + T_1 \left. \frac {\partial F_m^\beta}{\partial T} \right)_{V_m^{\beta}=V} (T_1) \nonumber \\
\nonumber \\
& & + \left[ \Delta_{t} F_m (T) + T \Delta_{PT} V_m (T) \frac{dp_c(T)}{dT} \right] \, \xi(T), 
\label{eqn17}
\end{eqnarray}

\noindent and the enthalpy 

\begin{eqnarray} 
\Delta H(T) &=& H_m^\beta(T) - H_m^\beta(T_1) + \Delta_{PT} H_m (T) \, \xi(T)  \nonumber \\
\nonumber \\
&=& \left[ U_m^\beta(T) + p_c(T) \, V_m^\beta(T) \right] - \left[ U_m^ \beta(T_1) + p_c(T_1) \, V_m^\beta(T_1) \right] \nonumber \\
\nonumber \\
& & + \left[ \Delta_{t} U_m (T) + p_c(T) \Delta_{PT} V_m (T) \right] \, \xi(T) \nonumber \\
\nonumber \\
&=& F_m^\beta(T) - F_m^\beta(T_1)- T \left. \frac {\partial F_m^\beta}{\partial T} \right)_{V_m^{\beta}=V} (T) + T_1 \left. \frac{\partial F_m^\beta}{\partial T} \right )_{V_m^{\beta}=V} (T_1) \nonumber \\
\nonumber \\ 
& & + p_c(T) \, V_m^\beta(T) - p_c(T_1) \, V_m^\beta(T_1) \nonumber \\
\nonumber \\
& & + T \frac{dp_c(T)}{dT} \Delta_{PT} V_m (T) \, \xi(T).
\label{eqn18}
\end{eqnarray}

\subsection{Constant volume heat capacity}

We have seen in Section \ref{sub1} that the constant pressure heat capacity exhibits a divergence during a first-order transformation at constant pressure. The analysis previously done on a constant volume transformation allows-us to examine what happens to the heat capacity at constant volume during the transition. Contrary to what happens in a constant pressure PT, in a constant volume PT we have shown that all the thermodynamic potentials and the entropy are continuous functions across the transformation. Therefore, no divergences are expected in this case. However, there can be discrete jumps in heat capacity, if different curvatures are met during the transformation.

Following equation (\ref{eqn9}), the heat capacity at constant volume of the whole system ($\beta$ and $\gamma$ phases) $C_V$ can be expressed as

\begin{equation} 
C_V = n^\beta c_V^\beta+ n^\gamma c_V^\gamma
\label{eqn19}
\end{equation}

\noindent where $c_V^i$ ($i=\beta,\gamma$) are the molar specific heats for each phase at constant total volume $V$.\footnote{$c_V^i$ takes into account not only the internal energy change due to a change of temperature, but also the one due to a variation of $n^i$ in the transformation (see Appendix A)} After some algebra (see Appendix A) we arrive at the following expression

\begin{eqnarray}
C_V &=& n^\beta \left[ c_{V_m^{\beta}}^\beta + T \, V_m^\beta \, \kappa^\beta \left( \frac {\alpha^\beta}{\kappa^\beta} - \frac{dp_c}{dT} \right)^2 \right] \nonumber \\
\nonumber \\
& & + \, n^\gamma \left[ c_{V_m^{\gamma}}^\gamma + T \, V_m^\gamma \, \kappa^\gamma \left( \frac {\alpha^\gamma}{\kappa^\gamma} - \frac{dp_c}{dT} \right)^2 \right]
\label{eqn20}
\end{eqnarray}

\noindent which is valid on the coexistence line. $c_{V_m^i}^i = \left. \frac{\partial U_m^i}{\partial T} \right)_{V_m^i}$ ($i=\beta,\gamma$) denotes the molar specific heat at constant molar volume of the $i$-phase.

We can evaluate this expression at the onset of the transformation, $T_1$, where $n \rightarrow n^\beta$ and $V \rightarrow V_m^{\beta}$. As $T \rightarrow T_1^+$, on the coexistence field, $\frac{dp_c}{dT} = \frac{\Delta_{PT} S_m}{\Delta_{PT} V_m}$. Thus, 

\begin{equation} 
\lim_{T\to T_i^{+}} C_V = n^\beta \left[ c_{V_m^{\beta}}^\beta + T \, V_m^\beta \, \kappa^\beta \left( \frac {\alpha^\beta}{\kappa^\beta} - \frac{\Delta_{PT} S_m}{\Delta_{PT} V_m}\right)^2 \right]
\label{eqn21}
\end{equation}

On the other hand, as $T \rightarrow T_1^-$, on the single-phase $\beta$ field, the limit is trivial,

\begin{equation} 
\lim_{T\to T_1^{-}} C_V = n^\beta c_{V_m^{\beta}}^\beta
\label{eqn22}
\end{equation}

\noindent It is clear from equations (\ref{eqn21}) and (\ref{eqn22}) that $C_V$ is discontinuous at $T_1$ (and at $T_2$, as a similar reasoning shows). Heat added to the system in the coexistence region not only is used to raise $T$ but to phase-transform as well. This is the origin of the discontinuity. 
During the constant $V$ transformation, as the system traverses the $p-T$ coexistence line, the other susceptibilities ($\alpha$, $\kappa$ and $C_p$) are strictly not defined (the limits from either side do not match) or can be thought of as divergences due to the finite volume or enthalpy changes without changes in pressure or temperature. 
Summarizing, during the constant $V$ transformation finite jumps are observed in $C_V$ only when entering and leaving the coexistence region, whereas $C_p$, $\alpha$ and $\kappa$ are not defined within the coexistence region. 

\section{Example: the liquid - ice VI transition in water}

\begin{figure}[h]
\centering
\includegraphics[width=\columnwidth]{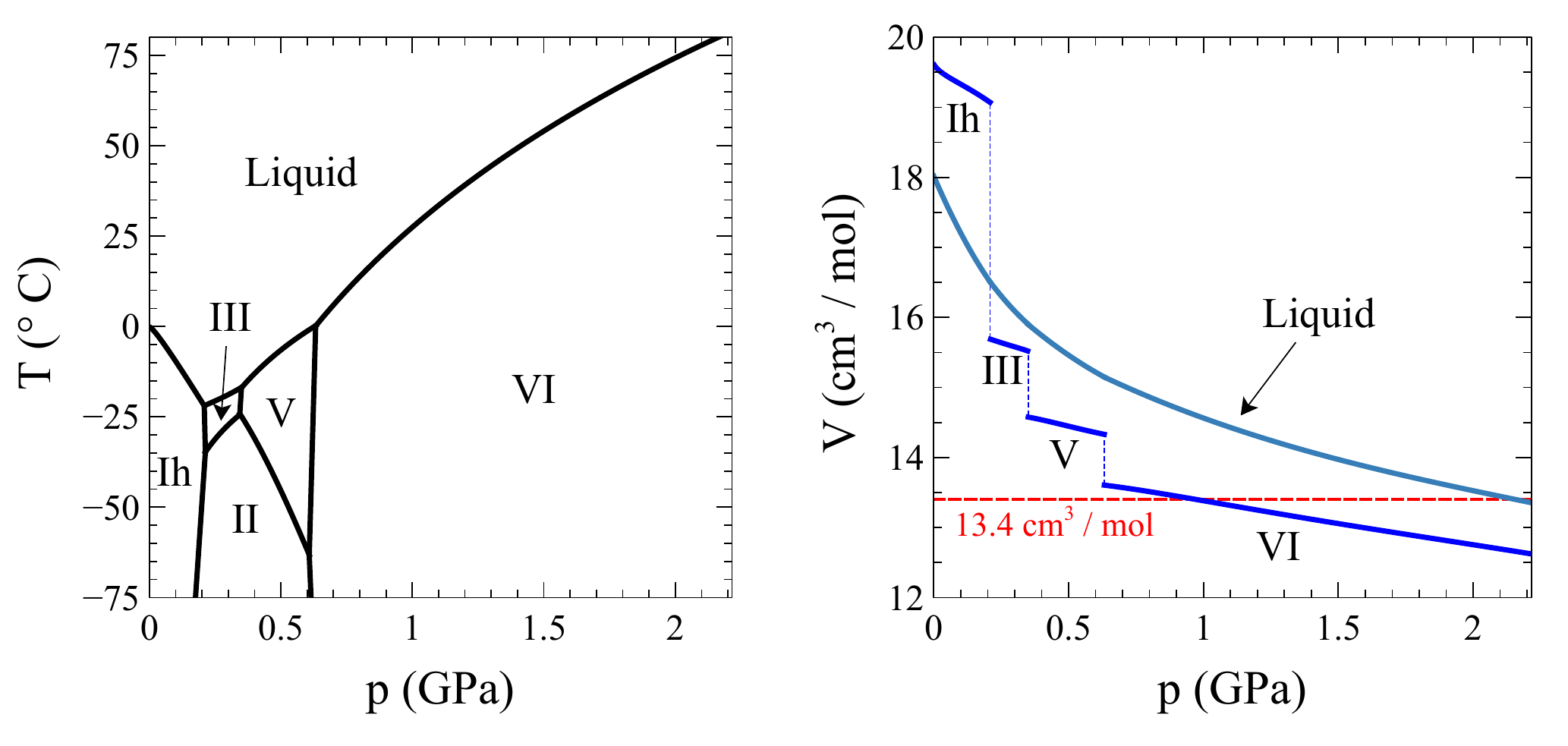}
\caption[]{Pressure-temperature phase diagram of water (left), and the molar volume of liquid and solid water along the phase coexistence line (right). Data taken from \cite{Duna}.}
\label{fig4}
\end{figure}

A \textit{sine qua non} condition for the possibility of a full transformation at constant volume is that the starting and final phases share the same molar volume. The bell-shaped coexistence region between vapor and liquid water in the $p-V$ diagram precludes this possibility. 
Figure \ref{fig4} shows the temperature vs. high-pressure phase diagram of water (left panel) together with a molar volume vs. pressure plot of the different ice phases and liquid water along the coexistence lines (right panel). 
It is clear from this figure that the only transition that meets the requirement of equal-initial-final volume is the liquid-ice VI one.

\begin{figure}[tbp]
\centering
\includegraphics[width=12cm]{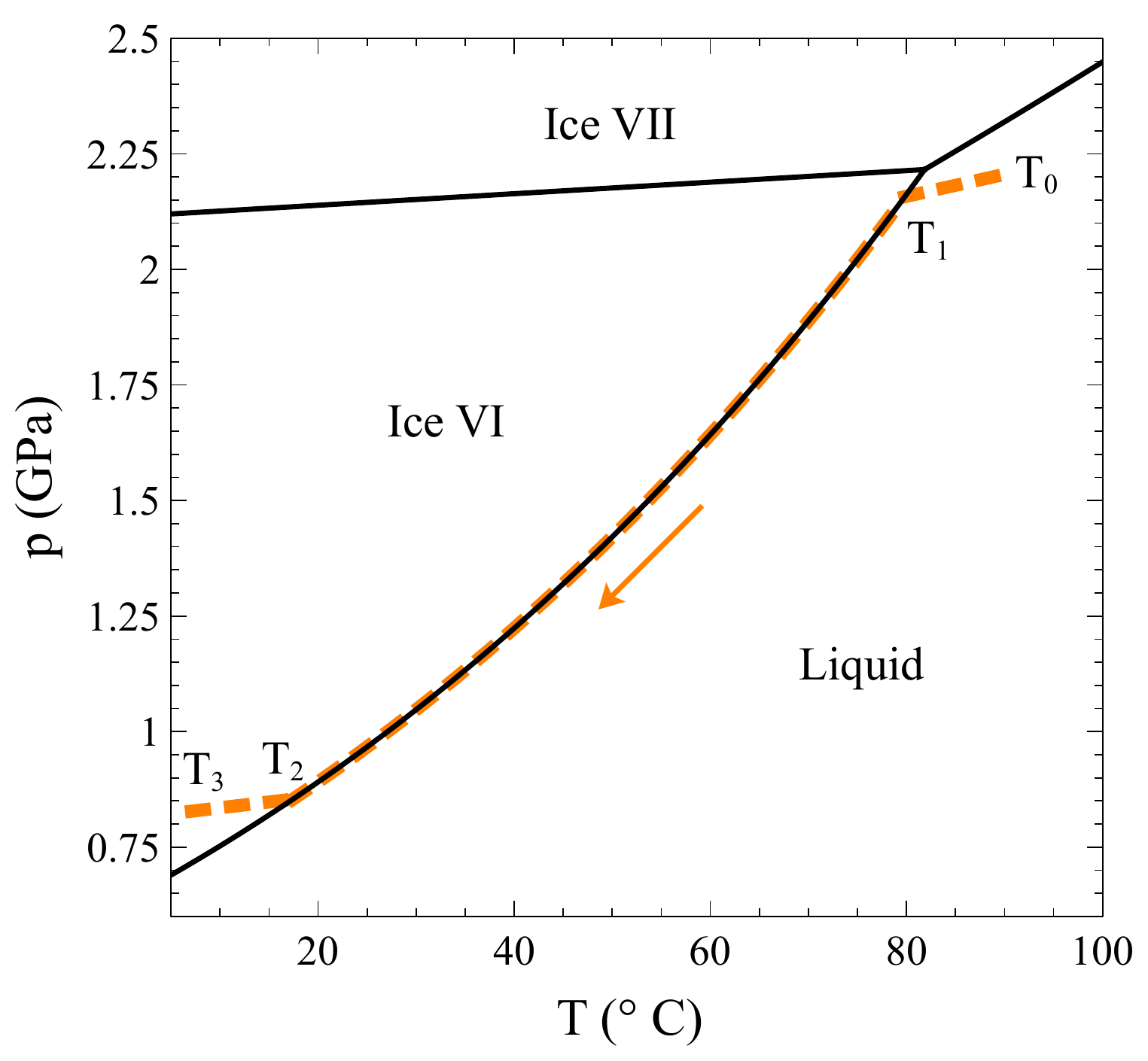}
\caption[]{Pressure-temperature phase diagram of water around the liquid - ice VI transition. The dashed line is the trajectory of a process at constant volume $V_0 =$ 13.4 cm$^3$/mol starting in the liquid phase ($T_0$) and ending up in the ice phase ($T_3$). The two phases coexist between $T_1$ and $T_2$.}
\label{fig5}
\end{figure}

So let us study this transition in detail. Using the complete thermodynamic information of liquid water \cite{liquid}, ice VI \cite{iceIV,Tchijov} and the coexistence line between both phases \cite{coexistence} we have calculated the trajectory of a process at constant volume $V_0 =$ 13.4 cm$^3$/mol that starts at $T_0$ = 90 $^o$C in the liquid phase, is cooled through the transition and ends up at $T_3$ = 7 $^o$C in the ice VI phase. The trajectory can be seen in the $p-T$ phase diagram of figure \ref{fig5}. Particularly, at $T_1$ = 79.81 $^o$C the liquid phase begins to transform into ice VI and finally at $T_2$ = 17.45 $^o$C the transformation is complete. 

\begin{figure}[p]
\centering
\includegraphics[width=12cm]{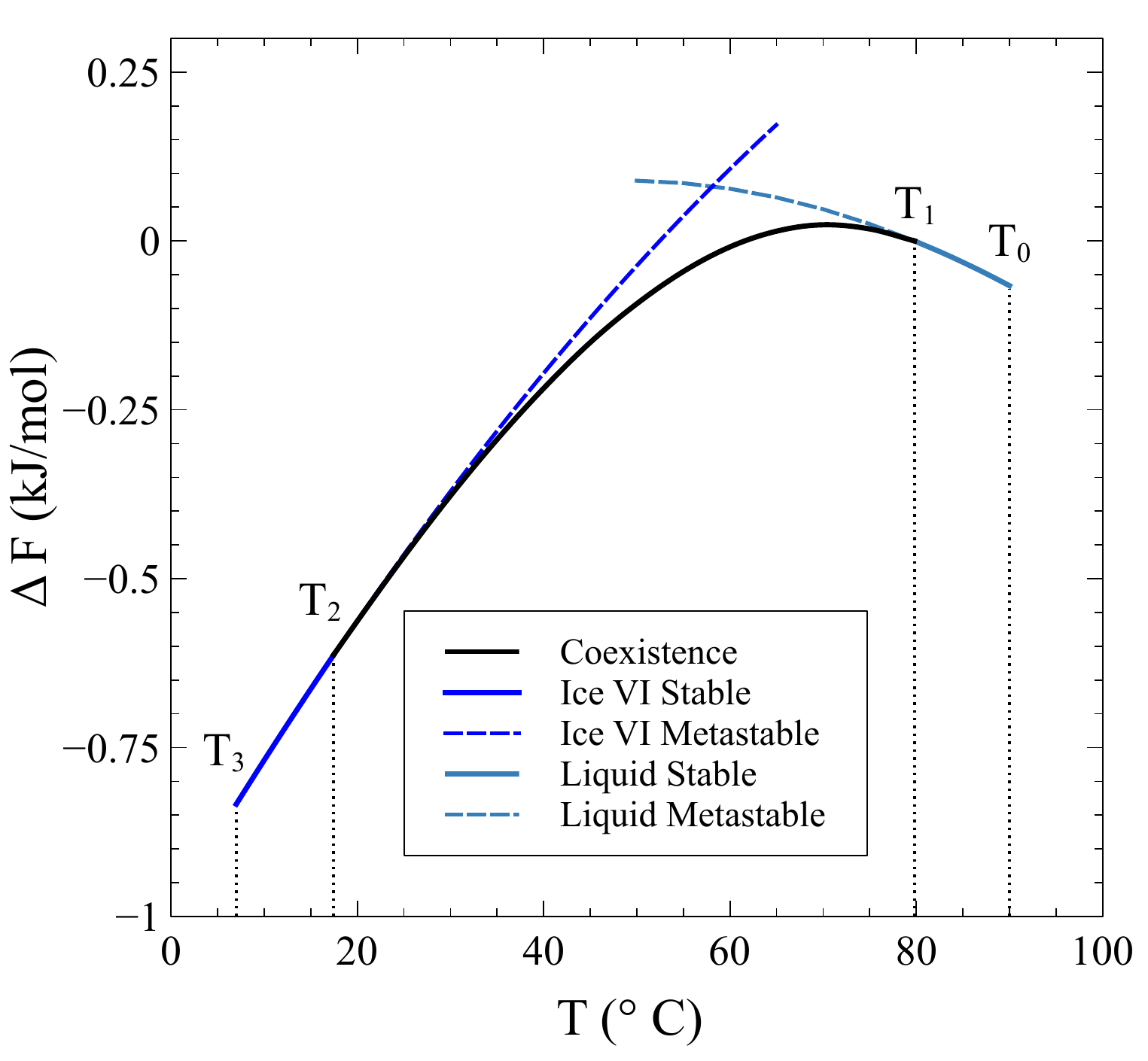}
\caption[]{The Helmholtz energy change $\Delta F(T) = F(T) - F_m^{liq} (T_1)$ in a process at constant volume $V_0 =$ 13.4 cm$^3$/mol. The dashed lines are the metastable parts of the pure-phase curves. The $F$ change for a prospective transformation at $p$ and $T$ (at each temperature) is shown for comparison.}
\label{fig6}
\end{figure}

The Helmholtz energy change $\Delta F(T) = F(T) - F_m^{liq} (T_1)$ along this constant volume path is displayed in Figure \ref{fig6}. The dashed lines correspond to the extrapolation of the single-phase curves into the coexistence region. 

First, it can be seen that $F$ values continuously change during the constant $V$ transformation. Secondly, it can be appreciated that the $F$ curve associated with the coexistence of $\beta$ and $\gamma$ phases always lies below the extrapolation of the single phase $F$ curves. This is the combination of the single phase $F$ curves previously mentioned in Section \ref{Vcte} that minimizes the Helmholtz energy at constant $V$ and at each $T$ in the coexistence region. 

\begin{figure}[p]
\centering
\includegraphics[width=12cm]{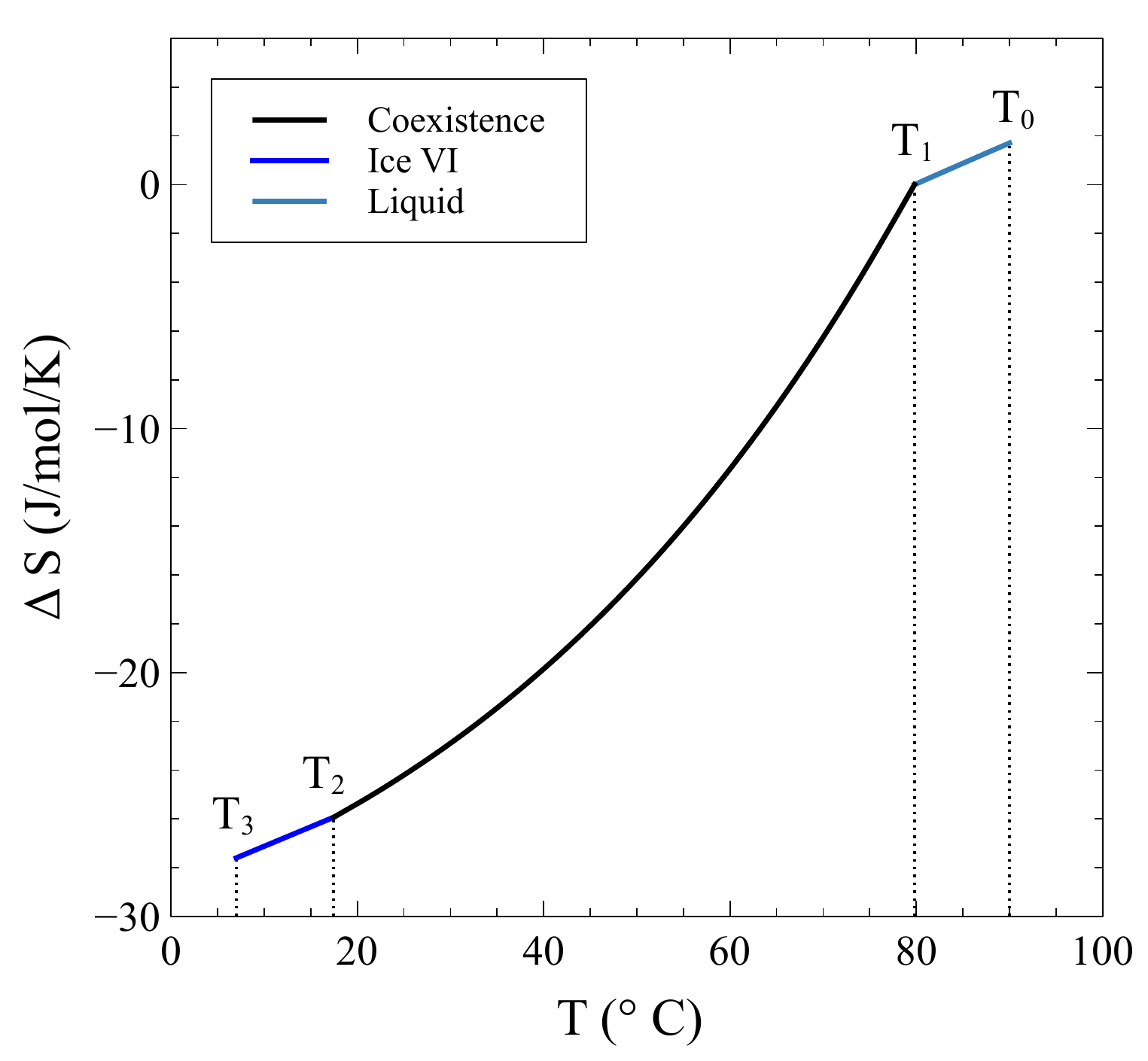}
\caption[]{The entropy change $\Delta S(T) = S(T) - S_m^{liq} (T_1)$ in a process at constant volume $V_0 =$ 13.4 cm$^3$/mol. The $S$ change for a prospective transformation at $p$ and $T$ (at each temperature) is shown for comparison.}
\label{fig7}
\end{figure}

Not only $\Delta F$ is continuous at $T_1$ and $T_2$, but its derivative is continuous too. This derivative is nothing but (minus) the entropy, which is shown in Figure \ref{fig7} as $\Delta S(T) = S(T) - S_m^{liq} (T_1)$. 

Its continuity during the transition has been previously mentioned and can be explicitly seen in this example. On the other hand, the derivative of the entropy, i.e. $C_V/T$, is clearly discontinuous at $T_1$ and $T_2$, as has been anticipated.

\begin{figure}[p]
\centering
\includegraphics[width=12cm]{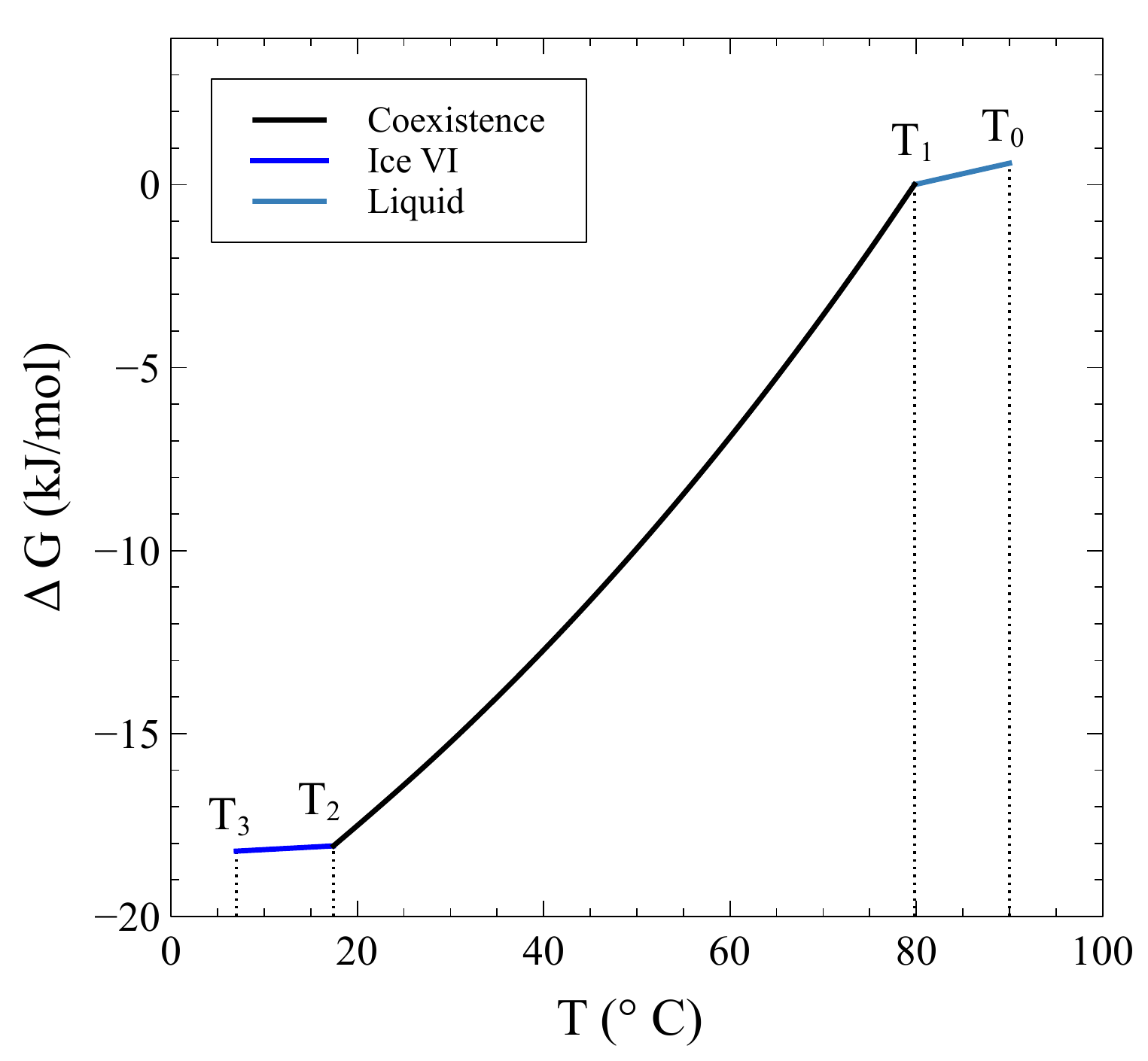}
\caption[]{The Gibbs energy change $\Delta G(T) = G(T) - G_m^{liq} (T_1)$ in a process at constant volume $V_0 =$ 13.4 cm$^3$/mol. The $G$ change for a prospective transformation at $p$ and $T$ (at each temperature) is shown for comparison.}
\label{fig8}
\end{figure}

\begin{figure}[p]
\centering
\includegraphics[width=12cm]{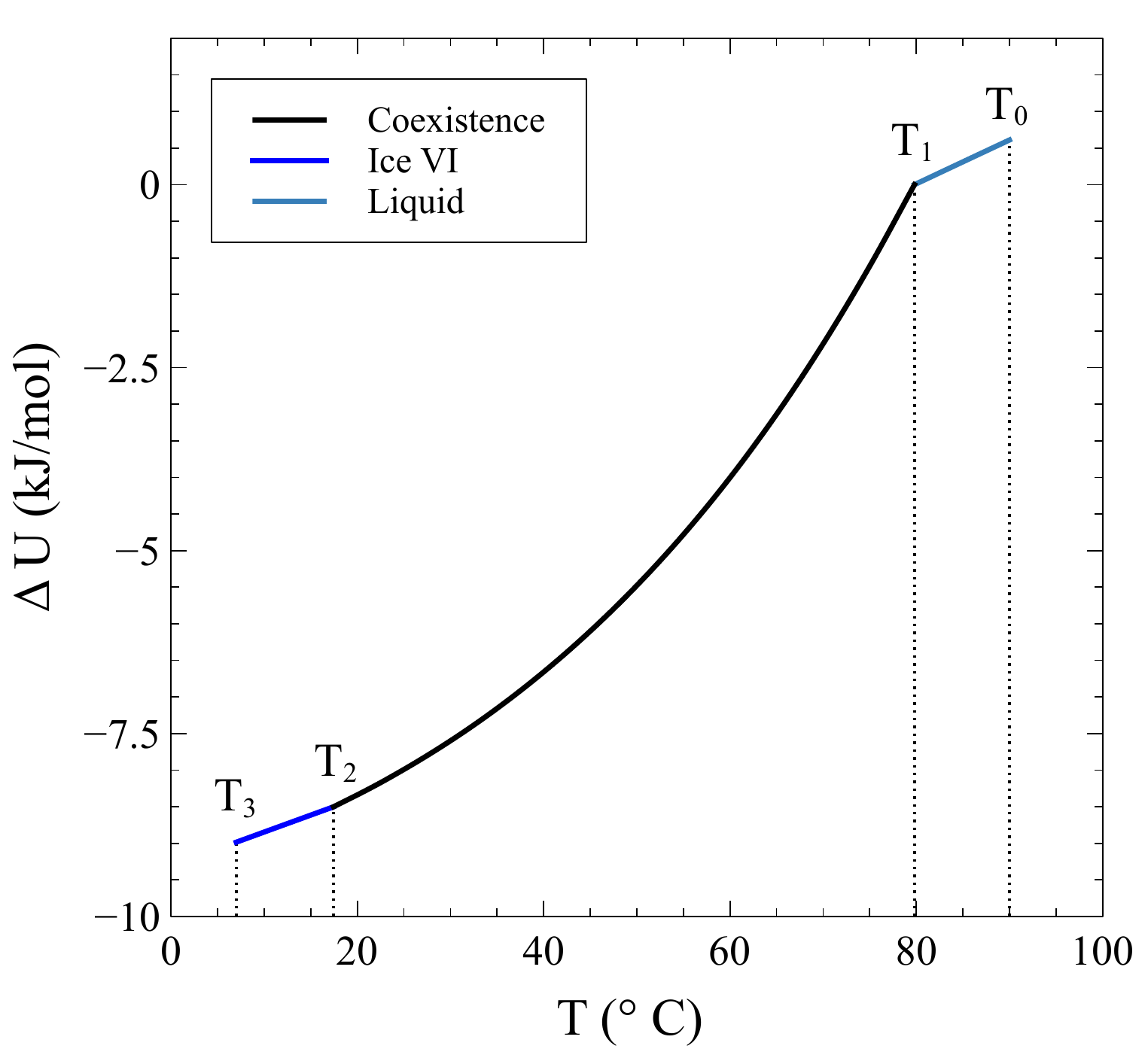}
\caption[]{The internal energy change $\Delta U(T) = U(T) - U_m^{liq} (T_1)$ in a process at constant volume $V_0 =$ 13.4 cm$^3$/mol. The $U$ change for a prospective transformation at $p$ and $T$ (at each temperature) is shown for comparison.}
\label{fig9}
\end{figure}

\begin{figure}[p]
\centering
\includegraphics[width=12cm]{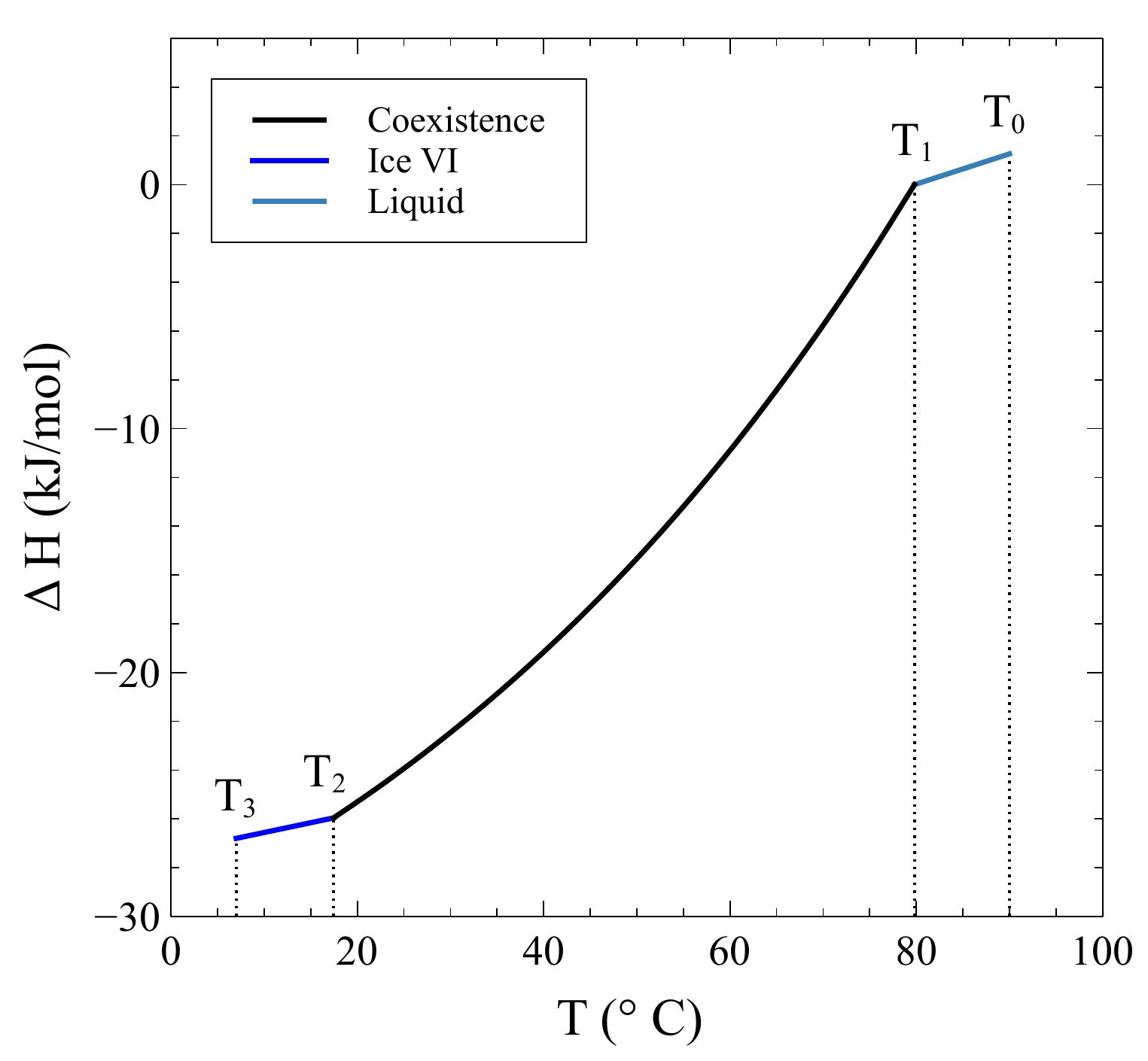}
\caption[]{The enthalpy change $\Delta H(T) = H(T) - H_m^{liq} (T_1)$ in a process at constant volume $V_0 =$ 13.4 cm$^3$/mol. The $H$ change for a prospective transformation at $p$ and $T$ (at each temperature) is shown for comparison.}
\label{fig10}
\end{figure}

Figures \ref{fig8}, \ref{fig9} and \ref{fig10} show the corresponding changes in the Gibbs energy $\Delta G(T) = G(T) - G_m^{liq} (T_1)$, the internal energy $\Delta U(T) = U(T) - U_m^{liq} (T_1)$and the enthalpy $\Delta H(T) = H(T) - H_m^{liq} (T_1)$, respectively. As predicted, all of them are continuous at $T_1$ and $T_2$. 

\begin{figure}[p]
\centering
\includegraphics[scale=0.7]{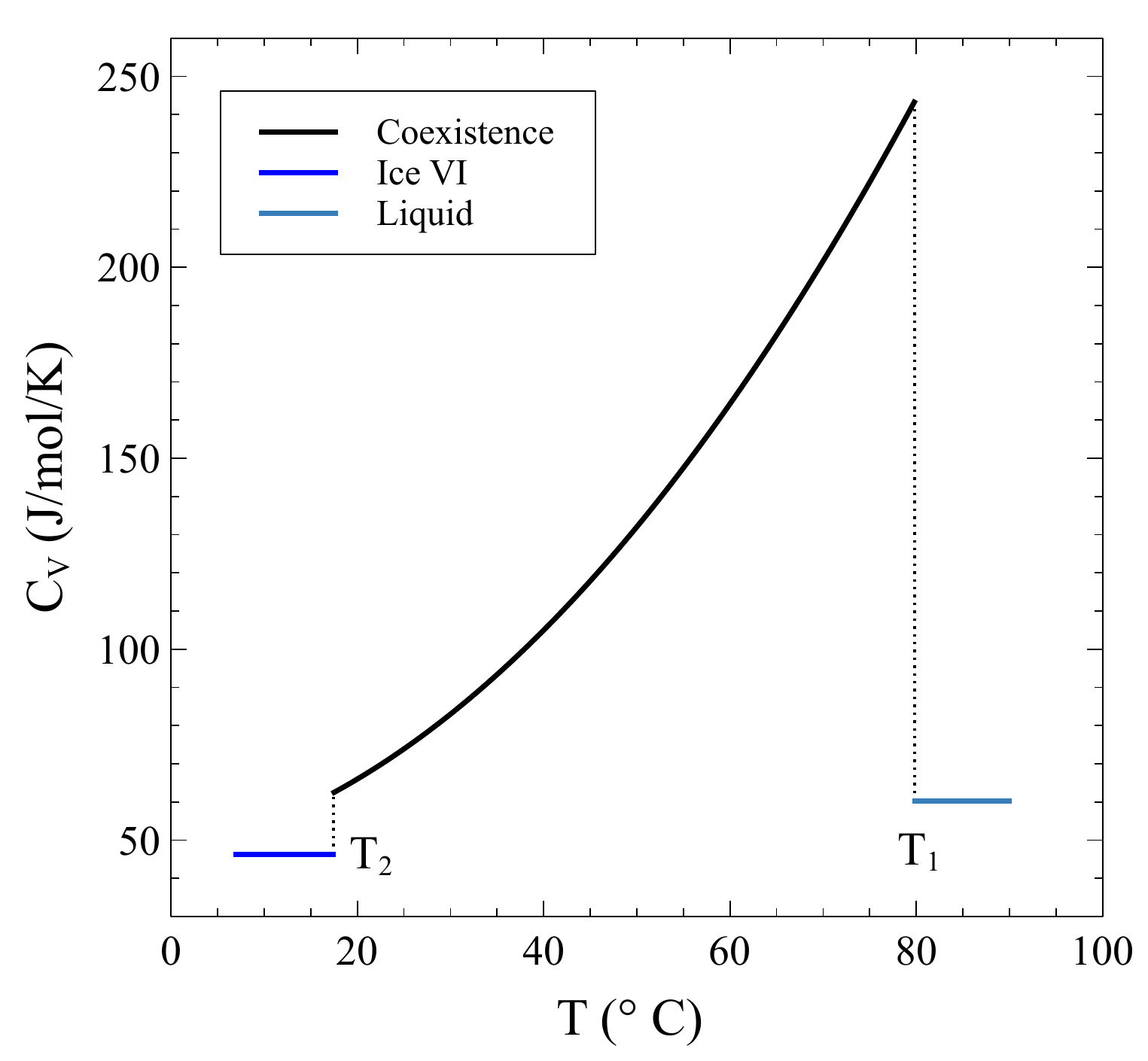}
\caption[]{Heat capacity at constant volume $V_0 =$ 13.4 cm$^3$/mol as a function of temperature. Jumps can be seen when entering and leaving the coexistence region.}
\label{fig11}
\end{figure}

Finally, Figure \ref{fig11} shows $C_V$ across the transition.

\section{Final remarks and conclusions}

Similarly to the constant volume process described here, it is also possible to go through a first-order phase transition in a constant entropy process. This is the basics of the so called mechanocaloric effect: the temperature change in an adiabatic system associated with the application of an external stress or pressure. The effect has been largely studied due to its potential applications in refrigeration technologies \cite{Manosa2017}. In any case, there is a full equivalence with the constant-V process. 

As conclusion, we want to remark that we have extensively described the thermodynamics of a constant volume phase transition for a simple system that exhibits a first-order phase transformation. With our approach we have shown that (i) the transformation extends over a finite range of $T$ (and $p$), (ii) each and every extensive potential (internal energy $U$, enthalpy $H$, Helmholtz energy $F$ and Gibbs energy $G$) and the entropy $S$ are continuous across the transition, unlike what is observed in a constant pressure (and temperature) transformation, and (iii) the constant volume heat capacity exhibits finite jumps when entering and leaving the two-phase coexistence region. 
These results have been illustrated by examining a first order phase transition in pure water, the transition at constant volume between liquid water and ice-VI. By using thermodynamic information available from the literature the transformation path in the $p-T$ plane and the characteristic $F$, $S$, $U$ and $C_V$ curves as a function of temperature have been calculated and discussed. 

We hope that these non-intuitive results, where a textbook first-order transition behaves as a continuous transformation, help students to get further insights into classical thermodynamics and phase transitions as-well.

\section{Acknowledgments}

The authors acknowledge Javier Luzuriaga for a careful reading of the manuscript. The authors are members of CONICET, Argentina. Work partially supported by SIIP 06/C559 and 06/C602 of Universidad Nacional de Cuyo, CONICET PIP2015-0610 and ANPCyT PICT2016-0204.

\appendix

\section{Constant volume heat capacity of the whole system}

Thermodynamic potentials are first-order homogeneous functions of their extensive variables. 
Hence, and according to Euler's theorem, the total internal energy of a pure substance when phases $\beta$ and $\gamma$ coexist in equilibrium can be written as

\begin{equation}
U(T,p, n^\beta, n^\gamma) = U_m^\beta[p,T] \, n^\beta + U_m^\gamma[p,T] \, n^\gamma
\label{A1}
\end{equation}

\noindent where $U_m^i$ is the molar energy of phase $i$ ($i = \beta, \gamma$). The constant volume specific heat of the whole system is

\begin{eqnarray} 
C_V &=& \left. \frac {\partial U}{\partial T} \right )_{V, n} \nonumber \\
\nonumber \\
&=& n^\beta \left. \frac {\partial U_m^\beta}{\partial T} \right )_{V, n} + 
n^\gamma \left. \frac {\partial U_m^\gamma}{\partial T} \right )_{V, n} +
U_m^\beta \left. \frac {\partial n^\beta}{\partial T} \right )_{V, n}+
U_m^\gamma \left. \frac {\partial n^\gamma}{\partial T} \right )_{V, n} \nonumber \\
\nonumber \\
&=& n^\beta \left. \frac {\partial U_m^\beta}{\partial T} \right )_{V, n} + 
n^\gamma \left. \frac {\partial U_m^\gamma}{\partial T} \right )_{V, n} +
(U_m^\beta - U_m^\gamma) \left. \frac {\partial n^\beta}{\partial T} \right )_{V, n}
\label{A2}
\end{eqnarray}

\noindent since the total mole number $n = n^\beta + n^\gamma$ is fixed. The derivative of the single-phase molar energy reads

\

\begin{eqnarray} 
\left. \frac {\partial U_m^i}{\partial T} \right )_{V, n} &=&
 \left. \frac {\partial U_m^i}{\partial T} \right )_p +
  \left. \frac {\partial U_m^i}{\partial p} \right )_T
   \left. \frac {\partial p}{\partial T} \right )_{V, n} \nonumber \\
   \nonumber \\
   &=&c_p^i-p V_m^i \alpha^i+V_m^i (\kappa^i p- \alpha^i T ) \left. \frac {\partial p}{\partial T} \right )_{V, n}
\label{A3}
\end{eqnarray}

\noindent where $c_p^i$, $\alpha^i = \frac {1} {V_m^i} \left. \frac{\partial V_m^i}{\partial T} \right)_{p}$ and $\kappa^i = -\frac {1} {V_m^i} \left. \frac{\partial V_m^i}{\partial p} \right)_{T}$ are the constant pressure specific heat, the thermal-expansion coefficient and the isothermal compressibility of the ith-phase, respectively. Additionally, $\left. \frac {\partial p}{\partial T} \right )_{V, n}=\frac{dp_c}{dT}$, as equilibrium requires. On the other hand, the conservation of the total volume 

\begin{equation} 
V_0=V(T,p, n^\beta, n^\gamma) = V_m^\beta[p,T] \, n^\beta + V_m^\gamma[p,T] \, n^\gamma
\label{A4}
\end{equation}

\noindent results in 

\begin{equation} 
\fl \left. \frac{\partial n^\beta}{\partial T} \right )_{V, n}  = 
\frac{1}{ V_m^\beta - V_m^\gamma } \left[ {-(n^\beta V_m^\beta \alpha^\beta + n^\gamma V_m^\gamma \alpha^\gamma) + (n^\beta V_m^\beta \kappa^\beta + n^\gamma V_m^\gamma \kappa^\gamma)
\frac {dp_c}{dT}}  \right ]  
\label{A5}
\end{equation}

\noindent Inserting (\ref{A3}) and (\ref{A5}) in (\ref{A2}) gives

\begin{eqnarray} 
\fl \left. \frac {\partial U}{\partial T} \right )_{V, n} =
\, n^\beta \left[ c_p^\beta-p V_m^\beta \alpha^\beta+V_m^\beta (\kappa^\beta p- \alpha^\beta T ) \frac {dp_c}{dT} \right] \nonumber \\
\nonumber \\
+ \, n^\beta \left [ c_p^\gamma-p V_m^\gamma \alpha^\gamma+V_m^\gamma (\kappa^\gamma p- \alpha^\gamma T ) \frac {dp_c}{dT} \right] \nonumber \\ 
\nonumber \\
+ \, \frac{(U_m^\beta - U_m^\gamma)}{ V_m^\beta - V_m^\gamma } \left[ {-(n^\beta V_m^\beta \alpha^\beta + n^\gamma V_m^\gamma \alpha^\gamma) + (n^\beta V_m^\beta \kappa^\beta + n^\gamma V_m^\gamma \kappa^\gamma)
\frac {dp_c}{dT}}  \right ]   
\label{A6}
\end{eqnarray}

\noindent The molar energy difference can be written as

\begin{equation} 
U_m^\beta-U_m^\gamma=G_m^\beta-G_m^\gamma - p( V_m^\beta-V_m^\gamma) + T( S_m^\beta-S_m^\gamma)
\label{A7}
\end{equation}

\noindent Phase coexistence at equilibrium requires $G_m^\beta = G_m^\gamma$ and the Clapeyron equation gives $ \frac{dp_c}{dT} = \frac{S_m^\alpha - S_m^\beta }{ V_m^\alpha - V_m^\beta }$. Hence, equation (\ref{A6}) can be rewritten as

\begin{eqnarray} 
\fl \left. \frac {\partial U}{\partial T} \right )_{V, n} = 
n^\beta \left[ c_p^\beta-V_m^\beta  \alpha^\beta T \, \frac{dp_c}{dT} \right] + 
n^\gamma \left [
c_p^\gamma- V_m^\gamma \alpha^\gamma T  \, \frac{dp_c}{dT} \right] \nonumber \\
\nonumber \\
+ \, T \, \frac{dp_c}{dT} 
\left[ -(n^\beta V_m^\beta \alpha^\beta + n^\gamma V_m^\gamma \alpha^\gamma) +
(n^\beta V_m^\beta \kappa^\beta + n^\gamma V_m^\gamma \kappa^\gamma)
\frac{dp_c}{dT}  \right]
\label{A8}
\end{eqnarray}

\noindent Since the constant pressure and constant volume specific heats are related by $c_p^i = c_{V_m^i}^i + \frac {T V_m^i {\alpha^i}^2} {\kappa^i}$, equation (\ref{A8}) becomes 

\begin{eqnarray} 
\fl \left. \frac {\partial U}{\partial T} \right )_{V, n} = 
n^\beta \left[  c_{V_m^i}^\beta +\frac {T V_m^\beta \alpha{^\beta}^2} {\kappa^\beta} - 2 V_m^\beta  \alpha^\beta T \, \frac{dp_c}{dT} + T V_m^\beta \kappa^\beta 
\left( \frac{dp_c}{dT} \right)^2 \right] \nonumber \\
\nonumber \\
+ \, n^\gamma \left [
c_{V_m^i}^\gamma +\frac {T V_m^\gamma \alpha{^\gamma}^2} {\kappa^\gamma} - 2 V_m^\gamma  \alpha^\gamma T \, \frac{dp_c}{dT} + T V_m^\gamma \kappa^\gamma 
\left( \frac{dp_c}{dT} \right)^2 \right]
\label{A9}
\end{eqnarray}

\noindent which finally can be conveniently rewritten to give 

\begin{eqnarray}
\fl \left. \frac {\partial U}{\partial T} \right )_{V, n} = n^\beta \left[ c_{V_m^{\beta}}^\beta + T \, V_m^\beta \, \kappa^\beta \left( \frac {\alpha^\beta}{\kappa^\beta} - \frac{dp_c}{dT} \right)^2 \right] \nonumber \\
\nonumber \\
+ \, n^\gamma \left[ c_{V_m^{\gamma}}^\gamma + T \, V_m^\gamma \, \kappa^\gamma \left( \frac {\alpha^\gamma}{\kappa^\gamma} - \frac{dp_c}{dT} \right)^2 \right]
\label{eqn26}
\end{eqnarray}

\end{document}